\documentclass[12pt]{article}
\usepackage{graphicx}
\usepackage{epsfig}
\usepackage{graphics}
\usepackage{amsmath, amsthm, amssymb}
\usepackage{color}

\newcommand\pubnumber{SLAC-PUB-12183}
\newcommand\pubdate{November 7, 2006}
\newcommand\hepnumber{hep-ph/0611091}

\def\SLAC{Stanford Linear Accelerator Center\\
  2575 Sand Hill Road, Menlo Park, California 94025 USA}
\def\doeack{\footnote{Work supported by the US Department of Energy,
                     contract DE--AC02--76SF00515.}}
\def\tev{\,{\ifmmode\mathrm {TeV}\else TeV\fi}}
\def\gev{\,{\ifmmode\mathrm {GeV}\else GeV\fi}}
\def\mev{\,{\ifmmode\mathrm {MeV}\else MeV\fi}}
\def\mpl{\ifmmode M_{pl}\else $M_{pl}$\fi}
\def\lsim{\mathrel{\rlap{\lower4pt\hbox{\hskip1pt$\sim$}}
    \raise1pt\hbox{$<$}}}                
\def\gsim{\mathrel{\rlap{\lower4pt\hbox{\hskip1pt$\sim$}}
    \raise1pt\hbox{$>$}}}                

\def\Title#1{\begin{center} {\Large #1 } \end{center}}
\def\Author#1{\begin{center}{ \sc #1} \end{center}}
\def\Address#1{\begin{center}{ \it #1} \end{center}}

\newcommand\pubblock{\rightline{\begin{tabular}{l} \pubnumber\\
         \pubdate \\ \hepnumber \end{tabular}}}

\begin{document}
  \begin{titlepage}
  \pubblock
  \vfill
  \Title{Microscopic Primordial Black Holes and Extra Dimensions}
  \vfill
  \Author{John Conley and Tommer Wizansky\doeack}
  \Address{\SLAC}
  \vfill
  \begin{abstract}
    We examine the production and evolution of microscopic black holes in the early universe 
    in the large extra dimensions scenario.  We demonstrate that, unlike in the standard
    four-dimensional cosmology, in large extra dimensions 
    absorption of matter from the primordial plasma by the black
    holes is significant and can lead to rapid growth of the black hole mass density.  This effect can 
    be used to constrain the conditions present in the very early universe. We demonstrate that this
    constraint is applicable in regions of parameter space not excluded by existing bounds.
  \end{abstract}
  \vfill
  \end{titlepage}
  \def\thefootnote{\fnsymbol{footnote}}
  \setcounter{footnote}{0}

  \section{Introduction}\label{sec:intro}
  The possible formation of black holes in the early universe
  has long been discussed.  The idea was first proposed by Carr and Hawking \cite{Carr:1974nx}, who
  considered the formation of such primordial black holes (PBHs) by the gravitational 
  collapse of density perturbations
  and their subsequent evolution.  They found that the PBH mass distribution is determined by the
  initial spectrum of density perturbations and the expansion of the universe, with accretion playing
  a negligible role. Soon
  after, Hawking discovered that black holes can emit particles \cite{Hawking:1974sw}, so that microscopic
  black holes decay very rapidly.
  PBHs smaller than about $10^{15}$ g would have evaporated by today, while larger ones could have
  survived.  It has been proposed that these relics could make up the cosmic dark matter \cite{Carr:1985tk}, 
  while on the other hand their non-observation constrains the initial spectrum
  of density fluctuations \cite{Carr:1975qj}.  It has also been suggested that the endpoint
  of black hole evaporation could be a stable Planck-sized remnant \cite{Bowick:1988xh}, 
  leading to additional observational consequences \cite{Barrow:1992hq}.

  Primordial black holes have also been considered in the context of extra dimensional theories 
  \cite{Argyres:1998qn}. In these theories, the fundamental scale of quantum gravity, called $M_*$, can 
  be as low as $1~\tev$. 
  It is well known that with extra dimensions, the properties 
  of microscopic black holes (those smaller than the size of the extra dimensions) are significantly 
  altered. A black hole in extra
  dimensions will be colder, larger, and longer lived than one of the same mass in four 
  dimensions \cite{Myers:1986un},
  with significant cosmological consequences.  In particular, production of PBHs by
  the collapse of primordial density perturbations in large extra dimensions has been studied
  \cite{Argyres:1998qn}.  The authors show that the unique properties of extra dimensional black
  holes lead to a relaxation of the bound on the spectral index.

  In this paper we discuss a different class of PBHs--microscopic black holes produced by high-energy 
  particle collisions in the early universe. The consequences of these tiny black holes
  have generally been neglected in the literature, under the premise that they are too hot and 
  short lived to have any observational effects.
  We argue that in the presence of extra dimensions,
  absorbtion of matter from the surrounding plasma cannot be neglected, and in fact can lead to 
  rapid growth.
  Consequently, the production and evolution of these black holes must be 
  analyzed. 
  As we will demonstrate, the mass density of 
  PBHs is determined by the initial
  temperature of the universe $T_I$, the number of extra dimensions $n$, and $M_*$. 
  We find that for different values of $n$, large regions of $T_I$--$M_*$ parameter space
  can be excluded by observational constraints.   It is worth noting that the effects of accretion
  in extra dimensional scenarios have previously been considered  
  \cite{Guedens:2002sd,Majumdar:2002mr,Tikhomirov:2005bt,Sendouda:2003dc,Sendouda:2004hz}.
  These authors analyzed rapidly growing black holes formed by other mechanisms
  in Randall-Sundrum cosmologies.

  There already exist constraints on theories of
  large extra dimensions \cite{Arkani-Hamed:1998nn,Hannestad:2003yd},  
  the most stringent of which come from astrophysical 
  considerations. For $n=1$ a natural value of 
  $M_*$ requires an extra dimension whose size is comparable to that of
  the solar system.  This would lead to obvious conlicts with observation.  
  For $n=2$ the overheating of neutron stars by captured Kaluza-Klein (KK) 
  gravitons constrains $M_*\gsim 700~\tev$.
  Larger values of $n$ are less tightly constrained.

  Independent bounds have already been placed by previous authors on the initial temperature of 
  the universe in these theories \cite{Hannestad:2001nq, Hall:1999mk}.  
  These are obtained by considering the production of KK gravitons in the early universe.
  The KK bounds can always be evaded by considering high enough values of $M_*$ and $n$, in which
  case the gravitons decay too early to have observational consequences.
  We demonstrate that in many cases these regions of parameter space are excluded by the PBH bounds.
  In addition, the graviton constraints can be evaded by, 
  for example, rapid graviton decay onto another brane or a heavier graviton
  spectrum arising from a complicated bulk geometry.  For example, Starkman, Stojkovic and Trodden \cite{Starkman:2000dy}
  have argued that all existing astrophysical and cosmological bounds can be evaded if the extra dimensions
  have the geometry of a compact hyperbolic manifold.
  The PBH constraints cannot be avoided so easily. 

  For definiteness, we consider the model of
  large extra dimensions proposed by Arkani-Hamed, Dimopoulos, and Dvali (ADD) 
  first presented in \cite{Arkani-Hamed:1998rs,Antoniadis:1998ig}.  In a subsequent paper \cite{Arkani-Hamed:1998nn} 
  these authors noted that bounds can be placed on the so-called ``normalcy'' temperature of the 
  universe---the temperature below
  which the extra-dimensional bulk must be stable in size and empty.  In contrast, we
  constrain the properties of the universe prior to attaining normalcy. 
  We do not specialize to any specific cosmological model.  Instead we consider 
  two possible thermal states for the extradimensional bulk without specifiying the 
  dynamics which lead to these states.  We examine scenarios where the bulk is cold and empty, 
  and where it is in thermal equilibrium with the brane.  
  
  The paper is organized as follows.  In Section~\ref{sec:evo} we analyze the evolution of a 
  single black hole in the hot primordial plasma.  In Section~\ref{sec:prod} we
  present a simple model for black hole production in the early universe, and use it to derive bounds
  on $T_I$ and $M_*$.  In Section~\ref{sec:asmp} 
  we show that the results we obtain with this simple model apply also to much less restrictive and more 
  realistic scenarios.
  Our conclusions are given in Section~\ref{sec:conc}.

  \section{Evolution of an extra-dimensional black hole}\label{sec:evo}

  In the ADD model, Standard Model particles are bound to a three-dimensional brane in a $3+n$-dimensional
  bulk space.
  The black holes relevant to this analysis are created on the brane and remain there. 
  The brane is populated by
  a thermal distribution of relativistic Standard Model particles.  We claim that, in this scenario,
  a black hole upon its creation will instantaneously (compared to the timescale of cosmological
  evolution) attain a maximum mass.  In the case that the bulk is in thermal equilibrium 
  with the brane, this mass
  will simply be the mass of a black hole the size of the extra dimension,
  \begin{equation}
    \label{eqn:ml}
    M_{max}=M_L=\left[\left(\frac{\mpl}{M_*}\right)^\frac{2}{n}\frac{1}{a_n}\right]^{n+1}M_*,
  \end{equation}
  where
  \begin{equation}
   \label{eqn:rn}
   a_n=\left(\frac{8\,\Gamma\left(\frac{n+3}{2}\right)}{(n+2)\pi^{\frac{n+1}{2}}}\right)^\frac{1}{n+1}.
  \end{equation}
  If the bulk is empty of energy density, the mass attained by the black hole depends on the temperature
  $T_0$ of the universe when the black hole is created and is given by
  \begin{equation}
    \label{eqn:mmax}
    M_{max}=\left(\gamma_n\frac{\mpl}{M_*^3}T_0^2\right)^\frac{n+1}{n-1}M_*,
  \end{equation}
  where 
  \begin{equation}
    \label{eqn:gamman}
    \gamma_n\equiv\sqrt{\frac{180}{\pi g_*}}\frac{n-1}{2(n+1)}\sigma_4 r_n^2\, .
  \end{equation}
  The remainder of this section is devoted to demonstrating this claim.

  The properties of black holes in
  infinitely large extra dimensions were first derived in \cite{Myers:1986un}.  In ADD, the extra
  dimensions are of finite size; the four dimensional Planck scale \mpl~and $M_*$ are related by 
  \begin{equation}
    \mpl^2\simeq L^n M_*^{n+2},
  \end{equation}
  where $L$ is the size of the extra dimensions.  We now review the properties of ADD black holes 
  as discussed in \cite{Argyres:1998qn}.  We note that they are only valid for black holes much smaller 
  than $L$. Larger black holes behave effectively four dimensionally.  
  The Schwartzschild radius
  of an ADD black hole of mass M is given by 
  \begin{equation}
    \label{eqn:rs}
    r_s=a_n\frac{1}{M_*}\left(\frac{M}{M_*}\right)^\frac{1}{n+1}.
  \end{equation}
  The temperature of the black hole is given by
  \begin{equation}
    T_{BH}=\frac{n+1}{4\pi r_s}.
  \end{equation}
  
  We consider a black hole with $r_s\ll L$ submerged in a thermal plasma at temperature $T$. If we ignore
  gravitational attraction, the absorption and emission can both be
  characterized by the Stefann-Boltzmann law.
  The rate of change of the black hole mass $M$ is then
  \begin{equation}
    \label{eq:dMdt}
    \frac{dM}{dt} = \sigma_4 A_4 (T^4-T_{BH}^4) + \sigma_{n+4} A_{n+4} (T^{n+4}-T_{BH}^{n+4}).
  \end{equation}
  Here, 
  \begin{eqnarray*}
    \sigma_4 & = & \frac{g_* \pi^2}{120}\;\;\;\;\mathrm{and} \\
    \sigma_{n+4} & = & \frac{g_b \Omega_{n+1} \Gamma(n+4) \zeta(n+4)}{2\pi^{n+3}(n+2)}
  \end{eqnarray*}
  are the 4 and $4+n$-dimensional Stefann-Boltzmann constants, and 
  $A_4$ and $A_{4+n}$ are the black hole surface areas on the brane and in the bulk.  
  The number of effective degrees of freedom on the brane is $g_*$ and $g_b=(n+1)(n+4)/2$ is
  the number of polarization states of a bulk graviton.

  In the early universe, the relationship between time, $t$, 
  and $T$ is determined by the 
  Friedmann equations.  We will show in the next section that the phase of black hole 
  growth takes place long before matter-radiation equality and that  
  the fraction of the universe's energy density
  in black holes is small.  Radiation domination can therefore be assumed, and
  \begin{equation}
    t=\sqrt{\frac{45}{16\pi^3 g_*}}\frac{\mpl}{T^2}.
  \end{equation}

\begin{figure}[t]
    \begin{center}
      \leavevmode
      \epsffile{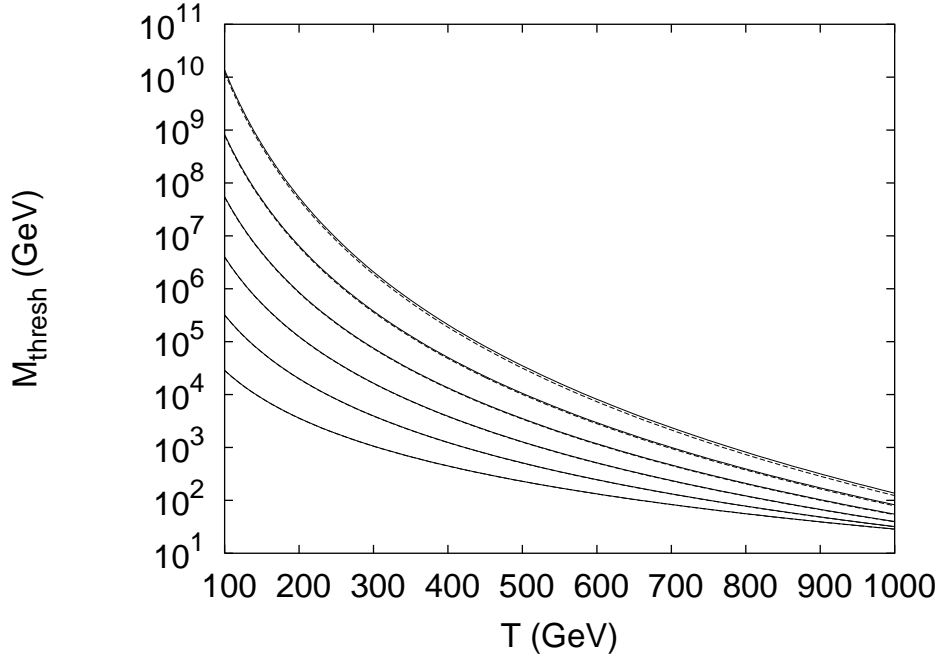}
    \end{center}
    \caption{$M_{thresh}$ (see text) as a function of temperature, for a thermalized (solid)
      and empty (dashed) bulk.  Here, $M_*=1~\tev$ and 
      from bottom to top, $n=2,3,4,5,6,7$.}\label{fig:mthresh}
  \end{figure}

  For a given temperature there is a threshold mass above which a black hole will absorb more than it emits.
  This mass, which we call $M_{thresh}$, is plotted in Figure~\ref{fig:mthresh}. It has almost the same value 
  whether the bulk is empty or thermalized.
  For \mbox{$M\gg M_{thresh}$}, we can neglect Hawking radiation entirely, and Equation~\ref{eq:dMdt}
  becomes
  \begin{equation}
    \label{dMdt_abs}
    \frac{dM}{dt} = \sigma_4 A_4 T^4 + \sigma_{n+4} A_{n+4} T^{n+4}.
  \end{equation}

  If the bulk is empty, the second term on the RHS can be dropped.
  The resulting equation can be trivially solved to obtain
  \begin{equation}
    \label{mass}
    M(T)=M_*\left[\left(\frac{M_0}{M_*}\right)^\frac{n-1}{n+1}+\gamma_n\frac{\mpl}{M_*^3}
      \left(T_0^2-T^2\right)\right]^\frac{n+1}{n-1},
  \end{equation}
  where $\gamma_n$ is given in Equation~\ref{eqn:gamman}.

\begin{figure}[t]
    \begin{center}
      \leavevmode
      \epsffile{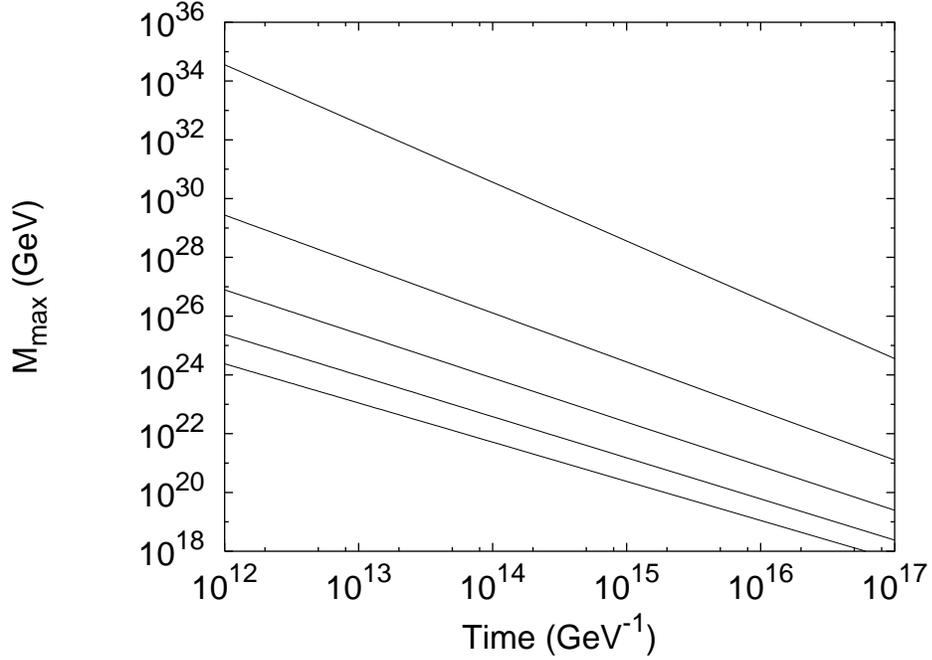}
    \end{center}
    \caption{The maximum mass $M_{max}$ attained by a black hole in the empty bulk scenario shown as a 
      function of time after the big bang.  From top to bottom, $n=3,4,5,6,7$.}\label{fig:branemass}
  \end{figure}

  This equation takes into account the competition between the growth of the black hole by
  the accretion of plasma and the cooling of the plasma by the expansion of the universe.
  But there is no competition.  
  Because $\mpl\gg M_*$, the second term dominates
  almost immediately and as the universe cools the mass rapidly approaches the value of $M_{max}$ 
  given in Equation~\ref{eqn:mmax}.  This value is plotted as a function of time for different values of
  $n$ in Figure~\ref{fig:branemass}.  Numerical integration confirms that, 
  in the regime considered, Hawking radiation is indeed negligible.

\begin{figure}[t]
    \begin{center}
      \leavevmode
      \epsffile{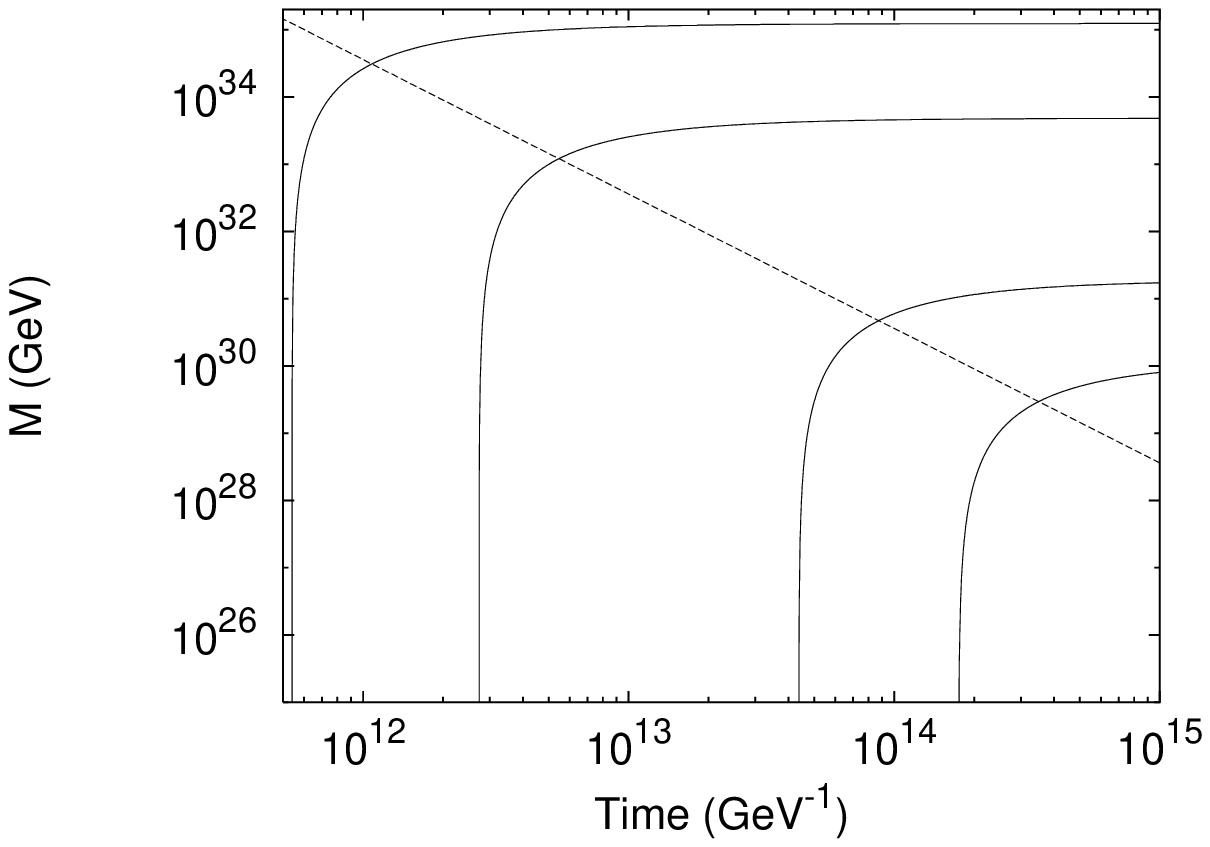}
    \end{center}
    \caption{The solid lines show the evolution of the mass of a black hole with $M_0=10M_{thresh}$ for 
      $M_*=1~\tev$ and $n=3$, assuming an empty bulk. From top to bottom, $T_0=1000,900,400,50~\gev$.  
      The dashed line is $M_{max}$.}\label{fig:growth} 
  \end{figure}

  In Figure~\ref{fig:growth}, the evolution of a black hole mass with an empty bulk is depicted.
  The almost instantaneous growth to $M_{max}$ is evident.
  In this plot, we take the initial mass $M_0=10M_{thresh}$.  We find, however, that
  the rapid growth depicted occurs even for initial masses extremely close to $M_{thresh}$.
  Similarly, initial masses even slightly less than $M_{thresh}$ lead to rapid decay 
  of the black hole.

  Now consider the opposite case in which the bulk is in thermal equilibrium with the brane.  
  For simplicity, consider integrating Equation~\ref{dMdt_abs} with the second term only.
  Once again the equation can be solved to obtain
  \begin{equation}
    M(T)=M_* \left[\gamma_n^{bulk}\frac{M_{pl}}{M_*^{n+3}}(T^{n+2}-T_0^{n+2})+
    \left(\frac{M_*}{M_0}\right)^{\frac{1}{n+1}}\right]^{-(n+1)},
  \end{equation}
  where
  $$
  \gamma_n^{bulk}\equiv\sqrt{\frac{45}{4 \pi^3 g_*}}\frac{\sigma_{n+4}\Omega_{n+2}a_n^{n+2}}{(n+1)(n+2)}\, .
  $$
  This solution will formally diverge after a finite time if
  \begin{equation}
    \label{Mdiv}
    M_0 > M_*\left(\frac{M_*^{n+3}}{\gamma_n^{bulk} M_{pl}T_0^{n+2}}\right)^{n+1} \equiv M_{div}.
  \end{equation}
  It should be noted that $M_{div}$ is
  typically much smaller than $M_{thresh}$.  Therefore the actual threshold for rapid growth of
  a black hole in the presence of a thermalized bulk is always given by $M_{thresh}$. 

  We can also ask
  how quickly the mass of the black hole grows.   The temperature of the universe corresponding 
  to the time when the black hole mass diverges is
  \begin{equation}
    \label{eq:Tdiv}
    T_{div}=T_0\left[1-\left(\frac{M_{div}}{M_0}\right)^{\frac{1}{(n+1)}}\right]^{\frac{1}{n+2}}.
  \end{equation}
  From an inspection of Equation~\ref{Mdiv} it is clear that for $M_0\gsim M_*$, $M_{div}\ll M_0$.
  Thus the second term in Equation \ref{eq:Tdiv} is much smaller than one and 
  the black hole mass diverges almost immediately.

  Of course, the mass of the black hole does not
  actually diverge.  In fact, once its size reaches that of the extra dimension the black hole
  begins behaving four-dimensionally and, as we will see below, the absorption
  effectively shuts off, so the black hole remains at $M_L$ as claimed. 
  
  We have seen that for both an empty and a thermalized bulk the evolution of a microscopic black
  hole in the early universe can be characterized by a threshold initial mass above which the PBH
  will rapidly grow and below which it will decay away.  In each case any rapidly growing 
  black hole reaches a uniform maximum mass $M_{max}$ which is shown in Table~\ref{tab:Mmax} 
  for different values of $n$ for both the empty and thermalized bulk.
  \begin{table}
    \begin{center}
      \begin{tabular}{|c|ccccc|} \hline
	$n$ & 3 & 4 & 5 & 6 & 7  \\
	\hline
	Empty & $1.9\times10^{35}$ & $1.1\times10^{30}$ 
	& $2.7\times10^{27}$ & $7.6\times10^{25}$ & $7.2\times10^{24}$ \\
	Thermalized & $1.4\times10^{46}$ & $3.9\times10^{43}$
	& $1.1\times10^{42}$ & $1.0\times10^{41}$ & $1.7\times10^{40}$ \\
	\hline
      \end{tabular}
    \end{center}
    \caption {The maximum mass, in $\gev$, of a rapidly growing black hole for $M_*=1 \tev$. The two rows correspond to an
      empty and thermalized bulk. For the empty bulk $T_0$ is taken to be $1$ TeV. For the 
      thermalized bulk the values are independent of temperature.}
    \label{tab:Mmax}
  \end{table}
  
  For four-dimensional black holes, Carr and Hawking showed \cite{Carr:1974nx} that one can neglect absorption.
  It is useful to review their argument and see why it does not apply for black holes in ADD. 
  In four dimensions the change in the black hole mass due to
  absorption is given by
  \begin{equation}
    \frac{dM}{dt} = \sigma_{4}(4\pi r_{s4}^2)T^4,
  \end{equation}
  where
  \begin{equation}
    r_{s4}=\frac{2M}{\mpl^2}.
  \end{equation}
  This has the solution
  \begin{equation}
    M(t\rightarrow\infty) = M_0\left[1-\left(\frac{720\sigma_4^2}{\pi g_*}\right)^{1/2}
      \frac{T_0^2M_0}{M_{pl}^3}\right]^{-1},
  \end{equation}
  which for black holes formed with sizes smaller than the horizon, never 
  gets very much larger than $M_0$. This is due to the large $M_{pl}$ suppression
  in the denominator. For
  this reason Carr and Hawking rightfully claimed that in four dimensions the absorption of particles 
  from the surrounding plasma can be safely neglected. It is the introduction of the low 
  mass scale, $M_*$, in theories with large extra dimensions which drastically
  alters the mass evolution of the PBHs.

  To complete this section we justify the thermodynamical treatment of black hole absorption.  This is valid
  so long as accretion takes place on a timescale much shorter than
  the lifetime of the black hole.  If this were not the case, the black hole could decay before a
  single particle collides with it.  For all $n>2$, we find that the lifetime of an $M_{thresh}$-sized black hole
  is much larger than the mean time between collisions with plasma particles.

  \section{Black hole production}\label{sec:prod}
\begin{figure} 
    \begin{center}
      \leavevmode
      \epsffile{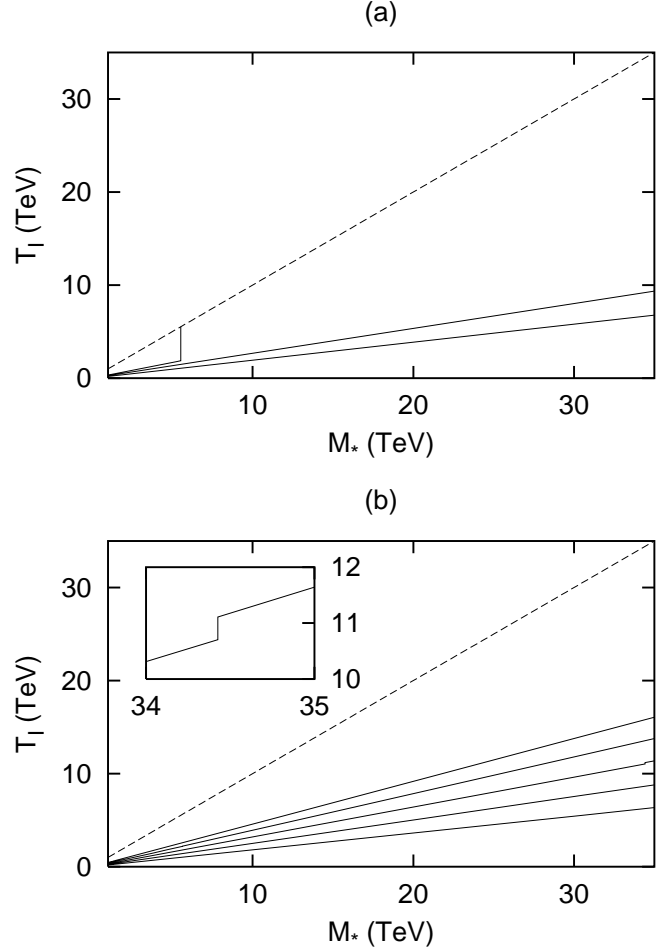}
    \end{center}
    \caption{Bounds on $T_I$ and $M_*$ from BBN and overclosure of the universe for an empty bulk. The 
      dashed lines represent $T=M_*$ above which the semi-classical description fails.
      The regions above the solid lines are ruled out. (a) For an empty bulk, from bottom to top, $n=3, 4, 5$.
      (b) For a thermalized bulk, from bottom to top $n=3, 4, 5, 6, 7$.  The inset is a magnification 
      of the $n=3$ curve showing the kink arising from a switch from the overclosure bound to the 
      BBN bound.}\label{fig:bound}
  \end{figure}

  We showed in the previous section that a black hole formed in the early universe with mass 
  above $M_{thresh}$ will immediately grow to mass $M_{max}$, while one with mass less than
  $M_{thresh}$ will decay away.  Combining this fact with the rate per unit volume
  for black hole production by particle collisions on the brane, we can easily evaluate the black hole mass 
  density produced in the early universe.  
  In this section we calculate this density and compare it to the critical density today 
  and to the radiation energy density during BBN to obtain bounds in the $T_I$--$M_*$ parameter space.  
  These bounds are shown in Figure~\ref{fig:bound} and constitute the main
  result of our paper.

  Define $\Gamma(t)$ to be the total rate per unit volume of black hole production.
  Then the black hole mass density at time $t$ obeys the equation
  \begin{equation}
    \label{eqn:y}
    \frac{d Y_{BH}(t)}{dt}=\frac{1}{s(t)}M_{max}(t)\Gamma(t),
  \end{equation}
  where $s$ is the entropy density of the universe, and 
  the black hole mass density $\rho_{BH}=sY_{BH}$. 
  For a thermalized bulk, $M_{max}$ is constant in time.  

  Here we have made two simplifications.  First, during production, we have taken
  the black hole mass density to be a small fraction of the total energy density
  of the universe.  Our bounds require that, at the very most, the black hole mass
  density $\rho_{BH}$ equals the radiation density $\rho_r$ at the time of big-bang nucleosynthesis
  (BBN). 
  As the universe cools, $\rho_{BH}\propto T^3$ and $\rho_r\propto T^4$. Thus, during black hole 
  production, which occurs at $T\gsim 100~\gev\gg T_{BBN}$, the mass density in 
  black holes is indeed negligible. Second,
  we ignore Hawking radiation during the production phase.  This is a good approximation as 
  the lifetimes of the black holes considered are much longer than the duration of the
  production phase.

  In a particle collision, the cross section for producing a black hole is 
  \cite{Banks:1999gd,Giddings:2001bu, Dimopoulos:2001hw}
  $\sigma(M)=f\pi r_s(M)^2$, where M is the invariant mass of the
  two-particle system, $r_s(M)$ is given by Equation \ref{eqn:rs} and $f$ is an order-one 
  constant.  
  For now we take $f=1$.
  If the brane fields are thermalized, the rate per unit volume $d\Gamma$ for creating
  black holes in the mass range $[M,M+dM]$ by particle collisions on the brane is
  \begin{equation}
    \label{eq:bhr}
    d\Gamma=dM g_*^2\int
    \frac{d^3k_1d^3k_2}{(2\pi)^6}f(\vec{k_1})f(\vec{k_2})\sigma
    (M)|\vec{v_1}-\vec{v_2}|
    \delta\left(\sqrt{(k_1^\mu+k_2^\mu)^2}-M\right),
  \end{equation}
  where $f(\vec{k})$ is the thermal distribution function.  Here we
  make the approximation that all species present in the universe are
  relativistic, and that for fermions and bosons alike we can use the
  Boltzmann distribution $f(\vec{k})=e^{-\frac{k}{T}}$.  At the temperatures
  we are considering, $T\sim 100-1000~\gev$, this is valid.

  We can do all but one of the integrals analytically, and we are left with
  \begin{equation}
    \frac{d\Gamma}{dM}=\frac{g_*^2a_n^2}{16\pi^3}T\left(\frac{M}{M_*}\right)^\frac{4+2n}{1+n}
    \int dk e^{-\frac{k}{T}}
    \left\{Me^{-\frac{M^2}{4kT}}-\sqrt{\pi kT}\left[\mathrm{Erf}\left(\frac{M}{2\sqrt{kT}}\right)-1\right]\right\}.
  \end{equation}
  To obtain the total rate of black hole production at a given
  temperature, we integrate over $k$ and $M$ numerically.  
  In order to take into account only the black holes that rapidly grow, and not those
  that decay away, the lower bound of the $M$ integration is set to be $M_{thresh}$.  One may worry
  that if $M_{thresh}\lsim M_*$, quantum gravitational effects may invalidate our calculation.  Fortunately,
  the bounds we will set restrict us to regions of parameter space for which $M_{thresh}$ is always
  significantly greater than $M_*$.

\begin{figure}[t]
    \begin{center}
      \leavevmode
      \epsffile{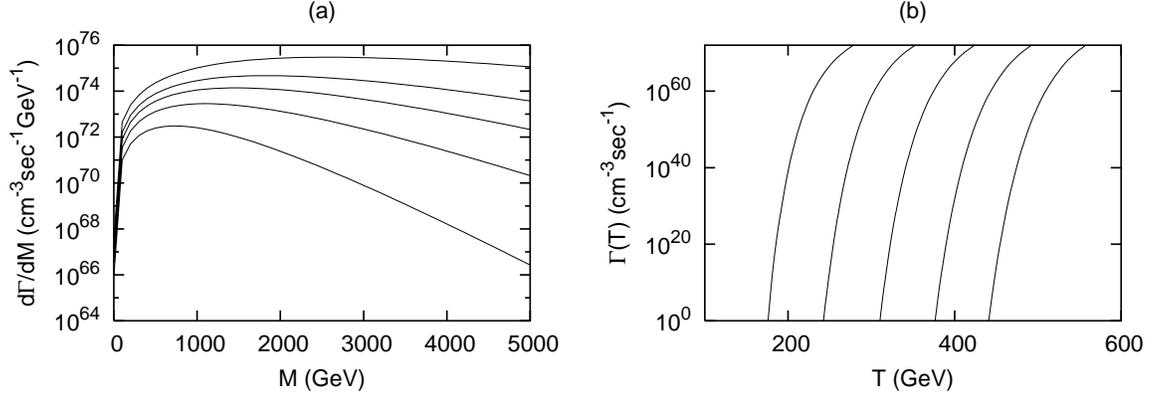}
    \end{center}
    \caption{(a) Differential production rate of microscopic black holes in the early
      universe for $n=3$. From top to bottom, $T= 700,500,400,300,200~\gev$. (b) Total 
      production rate. From left to right, $n=3,4,5,6,7.$}\label{fig:gamma}
  \end{figure}

  The differential production rate $d\Gamma/dM$ and total production rate $\Gamma$ are plotted in 
  Figure~\ref{fig:gamma}.  
  \epsfysize=5.5cm
  We see that the early universe is characterized by a period of intense black hole production, which
  falls off sharply with decreasing temperature.  It is also clear from this figure
  that the period of black hole production ends well before matter-radiation equality, as we
  assumed in the previous section.

  We solve Equation~\ref{eqn:y} to obtain $\rho_{BH}$ as a function of time. 
  We now include the effects of Hawking radiation by taking $\rho_{BH}=0$ at times
  later than the lifetime of the black holes produced.  For a thermal bulk, this is 
  just the lifetime of a black hole of mass $M_L$.  For an empty bulk, the vast majority
  of black holes are produced at temperatures very close to the initial temperature of the
  universe, $T_I$.  In that case we
  set $\rho_{BH}=0$ at times later than the lifetime of a black hole of mass $M_{max}(T_I)$.
  We can now impose the two previously mentioned constraints.  First, the black holes must
  not overclose the universe today.  Second, they must not form a significant fraction of the 
  energy density of the universe during big-bang nucleosynthesis. If either constraint were
  not satisfied, the expansion
  rate of the universe would be altered in a way that leads to measurable discrepancies
  with observation. Quantitatively, we require
  $$
  \frac{\rho_{BH}}{\rho_c}\Big |_{today}<1
  $$ 
  and  
  $$
  \frac{\rho_{BH}}{\rho_{r}}\Big |_{BBN}<1,
  $$
  where $\rho_c$ is the critical density.  Many other types of constraints on the primordial black hole
  abundance have been discussed.  For example, the decay of black holes today could alter the diffuse gamma
  ray spectrum in a measurable way \cite{Page:1976wx,Carr:1998fw,Halzen:1991uw}.  
  As we will see, however, the quantitative bounds on $T_I$ that we 
  derive are very insensitive to the nature of the observational constraint, so the two simple constraints  
  we use are sufficient.
  
  We use these conditions to bound the values of $T_I$ and $M_*$. In Figure~\ref{fig:bound} (a) and (b) 
  the region above the curves is ruled out for an empty and full bulk, respectively. 
  For an empty bulk the constraints are much weaker
  due to the lower value of $M_{max}$.  In fact, in this case, for all values of $n$ and $M_*$ considered,
  no black holes survive until today.  All bounds come from the BBN constraint, but for $n>5$ all
  black holes decay before BBN and no bound can be obtained.

  The $M_*$ dependence of the black hole lifetime accounts for the sharp cutoff on the bound for $n=5$ 
  with an
  empty bulk, above which the black holes decay before BBN.  It also accounts for the kinks in the bounds
  for $n=5$, $6$, and $7$ with a thermalized bulk, above which the black holes decay before today and the 
  overclosure bound is replaced
  by the BBN bound.  These kinks are barely visible; the two bounds are quantitatively
  almost identical.  This is because the black hole production rate is so incredibly sensitive to $T_I$, as
  can be seen in Figure~\ref{fig:gamma}(b).  As a result, a huge difference in black hole mass density can be
  acheived with a minute adjustment of $T_I$.  
  \epsfysize=13cm

  It is useful to compare these bounds to those obtained by Hannestad \cite{Hannestad:2003yd}
  and Hall and Smith \cite{Hall:1999mk} from
  gamma ray emission by the decay of KK gravitons.  For natural values of $M_*$ and low
  values of $n$, these bounds are much stronger than the PBH bounds.  Because \cite{Hannestad:2003yd}
  and \cite{Hall:1999mk} only consider $n=2,3$, we rederive their bounds for higher $n$ using
  Equation (12) in \cite{Hall:1999mk} and comparing to the COMPTEL data at $E=4~\mev$ as discussed
  in \cite{Hall:1999mk}.

\begin{figure} 
    \begin{center}
      \leavevmode
      \epsffile{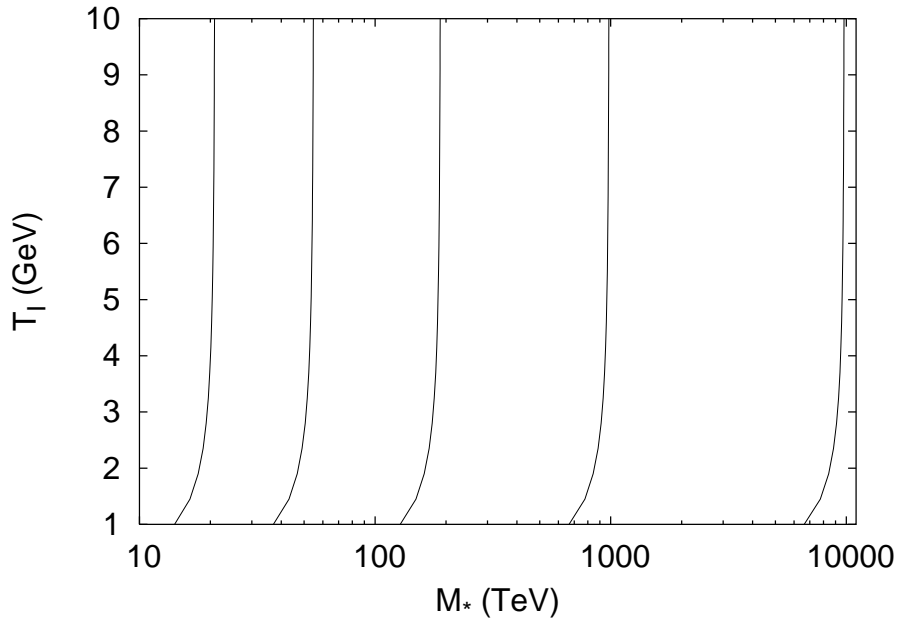}
    \end{center}
    \caption{Diffuse gamma ray bounds from KK graviton decay.  From left to right, $n=7,6,5,4,3$. The
      region to the left of the curves is excluded.}  
      \label{fig:KK}
  \end{figure}

  We find that the PBH bounds can be
  stronger in cases where the KK gravitons produced in the early universe decay
  before today and are not observed.  This occurs for higher values of $M_*$ and $n$.
  As can be seen in Figure \ref{fig:KK}, this results in a sharp cutoff of the KK bounds
  at a certain value of $M_*$ which rapidly decreases for increasing $n$.
  \epsfysize=8.5cm
  
  A comparison of Figure \ref{fig:KK} to Figure \ref{fig:bound}(b) shows that, for the full bulk case,
  for $n>5$ there are interesting regions of parameter space for which the PBH bounds dominate.  We note
  that a hot bulk in the very early universe may result in additional KK constraints, but a full analysis
  is beyond the scope of this paper.
  Comparing to Figure \ref{fig:bound}(a) shows that, for the empty bulk case, the KK bounds always dominate
  due to the decay of the black holes.  As we will see, however, if we allow for black hole remnants, 
  the cutoffs in the empty bulk PBH bounds go away.  In this case, for $n>5$ and 
  $M_*$ above the KK cutoff these become the dominant constraints.

  \section{Additional Considerations}\label{sec:asmp}
  
  In this section, we consider some possible extensions of our simple analysis.
  We first account for the fact that the black holes, which we took to be 
  Schwartzschild, are embedded in an expanding universe.  We then
  consider in more detail the uncertainties involved with realistic extradimensional
  black holes at the classical-quantum 
  threshold.  We show our conclusions to be robust and at most weakly dependent on the 
  above subtleties.  
  Finally, the consideration of black hole remnants leads to a 
  strengthening of our bounds.


  \subsection{Black holes in an FRW background}\label{subsec:frw}
  We have been somewhat simplistic in our analysis of black holes in the
  early universe.  In an expanding universe the usual Schwartzschild solution
  must be replaced by one which, asymptotically, is not flat but FRW.  Luckily,
  as noted by Carr and Hawking \cite{Carr:1974nx}, these corrections are only important
  for a black hole whose size approaches that of the horizon.  Since we are only
  considering temperatures lower than $M_*$ we have a lower bound on 
  the horizon size $R_h$ during the relevant epochs. We can compare 
  this bound to the maximum size the black holes attain, as derived in 
  Section~\ref{sec:evo}.

  With a thermalized bulk, the black holes reach the size of the extra 
  dimensions $L$.  For an empty bulk, the maximal black hole size is smaller.
  For $M_*=1~\tev$, $L/R_{h}(T=M_*)$ is shown in Table~\ref{tab:LoR}.
  We see that for all $n>2$ the black holes
  never approach the size of the horizon and we are thus well justified 
  in using the Schwartzschild solution.  For $n=2$, the black holes can grow 
  larger than $R_h$. 
  In this case our 
  analysis breaks down and a more careful study must be performed. But, because of the 
  existing stringent bounds on extra dimensional theories with $n=2$,
  we do not lose much by neglecting this case.

  \begin{table}[h!]
    \begin{center}
      \begin{tabular}{|c|cccccc|} \hline
	$n$ & 2 & 3 & 4 & 5 & 6 & 7  \\
	\hline
	$L/R_h$ & $17.15$ & $7.45\times10^{-5}$ & $1.55\times10^{-7}$ 
	& $3.82\times10^{-9}$ & $3.23\times10^{-10}$ & $5.54\times10^{-11}$ \\
	\hline
      \end{tabular}
    \end{center}
    \caption {The ratio of the size of the extra dimensions to the horizon size,
      for $T=M_*=1~\tev$.}
    \label{tab:LoR}
  \end{table}
 
  \subsection{Properties of extra dimensional black holes}\label{subsec:prop}
  As was noted in the previous sections, the exact properties of extra 
  dimensional black holes of mass close to the fundamental Planck scale are 
  rather poorly understood. Specifically, there is no consensus regarding the 
  production cross section or the exact spectrum of Hawking radiation.  So far,
  we have used the canonical choice of  $\sigma(M)=\pi r_s(M)^2$ and a purely 
  thermal spectrum.  In doing so, we have neglected the effects of angular 
  momentum and the dissipation of energy through gravitational waves in 
  particle collisions. We have also ignored radiative gray-body factors.

  As previously mentioned, the black hole production cross section is given by 
  $$
  \sigma(M)=f\pi r_s(M)^2.
  $$
  The factor $f$ depends on the center of mass energy, the orbital angular 
  momentum and the spin of the interacting particles (see \cite{Kanti:2004nr} and
  references therein). Typical values for $f$
  can reach as low as $\sim0.5$. In addition, the total energy
  actually trapped behind the black hole event horizon is not necessarily $\sqrt{s}$,
  as we have assumed.  Studies have concluded that only approximately $40\%-90\%$ 
  of the collision center of mass energy actually forms the black hole, with the 
  rest escaping in the form of gravitational radiation.  Also, a black hole
  created with a high initial angular momentum will undergo a rapid ``spin-down''
  phase during which it will shed its angular momentum and a significant
  fraction of its mass \cite{Ida:2006tf}.  This effect could result in an even lower
  initial mass of the Schwartzschild black hole.
  
  As we have seen, however, the
  black hole mass density is exquisitely sensitive to the initial temperature.
  An order one modification of the cross section will thus have a negligible
  effect on the bounds on $T_I$. The percentage of energy converted to 
  black hole mass will affect the number of such black holes which are created 
  above the threshold for rapid growth, and thus the final PBH mass density.  But,
  once again, a tiny modification of the intial temperature would compensate 
  for this effect, leaving the final bounds essentially unchanged. The correction to the 
  spectrum of Hawking radiation due to gray-body factors, because it could modify
  the black hole lifetime, could change the location of the kinks and cutoffs in
  Figure~\ref{fig:bound}, but would not qualitatively alter our conclusions.

  \subsection{Black hole remnants}
  Many authors have explored the possibility that black holes do not evaporate away completely but 
  leave behind a microscopic remnant.  This remnant is typically of the fundamental mass, which 
  in large extra dimensions is $M_*$.  Black hole remnants have been proposed in various 
  contexts.  Adler, Chen and Santiago \cite{Adler:2001vs} argued that the quantum mechanical 
  uncertainty principle must 
  be generalized in the presence of a curved space-time, causing Hawking radiation to shut off once the 
  black hole reaches the fundamental scale.  Rizzo \cite{Rizzo:2005fz} investigated the 
  thermodynamics of black
  holes in the presence of higher curvature gravity and found that the specific heat
  of these black holes can become positive.  The black hole would then cool as it evaporates, 
  asymptoting to a finite size.

  We have incorporated the possibility of black hole remnants by extending the bounds of 
  integration of Equation \ref{eq:bhr}.  Originally we integrated over mass values between 
  $M_{thresh}$ and $\infty$ while assuming that, below $M_{thresh}$, the black holes would 
  decay away.  With remnants, we should 
  extend the region of integration down to the quantum gravity limit, taken to be several times 
  the fundamental scale. To each black hole created below 
  $M_{thresh}$ we assign a final mass $M_*$.  The resulting bounds on $T_I$ are approximately 
  $1-2$ orders of magnitude lower than those displayed in Figure~\ref{fig:bound} and cover the entire
  range in $M_*$ for all values of $n$ for both a full and empty bulk.
  The bounds are sensitive to the value of the remnant mass.  Lowering 
  this mass below $M_*$ can substantially
  strengthen the bounds on $T_I$.

  \section{Conclusions}\label{sec:conc}
  In this paper we have presented a new constraint on cosmological models in
  theories of large extra dimensions, stemming from the production of rapidly
  growing microscopic black holes in the very early universe. We found 
  that generic upper bounds can be placed
  on the temperature of the universe in any post inflationary epoch.  As an example 
  we analyzed two simple cases, that of a completely empty bulk and a bulk that is 
  fully thermalized with the standard model brane. We found that in both cases
  significant regions of $T_I-M_*$ parameter space can be excluded.  Notably, these
  bounds are not sensitive to many of the details relating to  black holes at the 
  classical-quantum 
  threshold.  This eliminates one of the main sources of uncertainty which plague 
  typical studies of black holes in extra dimensions.

  When compared to existing bounds, we showed that the PBH constraints are stronger for high
  values of $n$ and $M_*$.
  Moreover, in the previously mentioned scenarios where the bulk is depopulated,
  the graviton bounds are weakened even further.
  On the other hand, at low values of these parameters, our bounds are
  generally weaker.
  This is because while the cross section for black hole production is large,
  only sufficiently massive ones are long-lived.  At low temperatures, these represent
  a small fraction of the black holes produced.  Even if one includes the possibility
  of remnants, the black hole mass is bounded from below by the quantum gravity 
  threshold.  For temperatures much lower than $M_*$ very few will be produced above this
  limit.  

  Although we focused on specific examples, it is important to note that the 
  phenomenon of primordial black hole production and subsequent growth is generic.
  In any theory of extra dimensions where the fundamental Planck scale is low,
  a large enough energy density will lead to this effect.  Generalizations of our
  simple scenario could include for example non-thermal brane particle distributions, 
  a range of bulk thermal states, and a non-static bulk radius.  Rapidly growing
  microscopic black holes in the early universe represent a conceptually new phenomenon which
  should be considered in detail.

  \section{Acknowledgements}
  The authors would like to thank Mustafa Amin, Pisin Chen, Savas Dimopoulos, JoAnne Hewett, 
  Andre Linde, Michael Peskin, 
  Tom Rizzo, and Jay Wacker for very helpful discussions.
  This work is supported by the US Department of Energy, contract DE--AC02--76SF00515.
  

\end{document}